\documentclass{PoS}

\title{Quantum Mechanics \`a la Langevin and Supersymmetry}

\ShortTitle{Quantum Mechanics \& SUSY}

\author{\speaker{Stam Nicolis}\\
        CNRS--Laboratoire de Math\'ematiques et Physique Th\'eorique (UMR7350)\\
        F\'ed\'eration de Recherche ``Denis Poisson'' (FR2964)\\
        D\'epartement de Physique, Universit\'e``Fran\c{c}ois Rabelais'' de Tours\\
        Parc Grandmont, 37200 Tours, France\\
        E-mail: \email{Stam.Nicolis@lmpt.univ-tours.fr}}

\abstract{
      We study quantum mechanics in the stochastic formulation, using the functional integral approach. The noise term enters the classical action as a local contribution of anticommuting fields. The partition function is not  invariant under ${\mathcal N}=1$ SUSY, but can be obtained from a, manifestly, supersymmetric expression, upon fixing a local fermionic symmetry, called $\kappa-$symmetry.
       The kinetic term for the fermions is a total derivative and can contribute only on the boundaries. 
     We define combinations that scale appropriately, as the lattice spacing is taken to zero and the lattice size to infinity and provide evidence, by numerical simulations, that the correlation functions of the auxiliary field do satisfy Wick's theorem. We show, in particular, that simulations can be carried out using a purely bosonic action.    
 
 The physical import is that the classical trajectory, $\phi(\tau)$, becomes a (chiral) superfield, $(\phi(\tau),\psi_\alpha(\tau),F(\tau))$, when quantum fluctuations are taken into account.    
          
          }

\FullConference{31st International Symposium on Lattice Field Theory - LATTICE 2013\\
		July 29 - August 3, 2013\\
		Mainz, Germany}

\begin{document}

\section{Introduction}
Supersymmetry is a symmetry that unifies internal and space-time symmetries. Within the context of quantum field theory it implies the existence of particles, ``partners'' to the known particles. The absence of such partners, e.g. a spin-0 particle with the same charge and mass as the electron, implies that supersymmetry must be broken at the energy scales that have been explored so far. How it is broken, within the framework of the Standard Model,  is unknown.
It is, therefore, useful to explore this issue within a framework that is not sensitive to perturbation theory. Lattice field theory is such a framework. 

However there are challenges, technical and conceptual, in carrying out this program~\cite{catterall}. On the other hand, supersymmetry, as a symmetry of space-time,  should have consequences beyond particle physics. To date the driving force behind supersymmetry is that it provides the simplest way of stabilizing the hierarchy between the weak scale and the Planck scale and controlling the corrections to the mass of the Higgs boson. This still leaves a large parameter space to explore. 
An obstacle in our understanding of the role of supersymmetry is that we do not know how to explore ``superspace'', where supersymmetry acts ``naturally''.

In an effort to acquire some intuition about how supersymmetry may be realized and broken, it might be helpful to study it in  ``simpler'' situations. For the intuition so acquired to be useful, it is necessary that the simplicity not eliminate the signal. In the Standard Model, the supersymmetry, that is relevant, is {\em target space} supersymmetry: the particles are target space fermions or bosons, regarding statistics and are described by spinors, 
vectors--and scalars--regarding spin; so are their putative superpartners. In less than four space-time dimensions, where the rotation group is abelian, we may talk of commuting or anticommuting fields; but the statistics is not, necessarily, that of fermions or bosons. And it might not be possible to define spinors. 

A conceptual stumbling block, in these situations,  indeed, is,  how we can define supersymmetry.  The definition we shall use is the following: that the action is invariant under transformations, whose parameters are anticommuting variables {\em and} the anticommutator of two such transformations closes on the generator of the translations. 

The situation that we will study is quantum mechanics, as a quantum field theory in one space-time dimension. There have been extensive studies of ``supersymmetric quantum mechanics''~\cite{SUSYQM}, however these used a canonical formalism--i.e. worked in phase space, rather than configuration space and the aim was rather to understand how supersymmetric field theoretic notions appeared in the simpler setting.  
Lattice investigations~\cite{latticeSUSYQM} have, also, focused in this direction. 

So it is useful to try to set up the framework for the opposite direction. In this way, in fact, we can study superspace.

The starting point of our investigation is the stochastic formulation~\cite{stochastic}. In this  formulation we write the Langevin equation
\begin{equation}
\label{langevin1}
\frac{\partial\phi}{\partial t}=-\frac{\partial U}{\partial\phi} +\eta(t)
\end{equation}
where $t$ is the equilibration time--we are interested in the limit $t\to\infty$. The field, $\eta(t)$, is a Gaussian stochastic process:
\begin{equation}
\label{gaussian_noise}
\begin{array}{l}
\left\langle\eta(tà\right\rangle = 0\\
\left\langle\eta(t)\eta(t')\right\rangle = \delta(t-t')\\
\left\langle
\eta(t_1)\eta(t_2)\cdots\eta(t_{2n})
\right\rangle
= \sum_{\pi}
\left\langle \eta(t_{\pi(1)})\eta(t_{\pi(2)})\right\rangle\cdots
\left\langle \eta(t_{\pi(2n-1)})\eta(t_{\pi(2n)})\right\rangle
\end{array}
\end{equation}
where the sum is over all perumutations. 

At equilibrium, $t\to\infty$,
\begin{equation}
\eta = \frac{\partial U(\phi)}{\partial\phi}
\end{equation}
If the fields, $\eta$ and $\phi$, are random variables, we have a problem in probability theory, whose solution can provide interesting insights for physics~\cite{nicolis_zerkak}. If they are functions of other variables, then we have a problem in quantum field theory.

If $U(\phi)$ is a local functional of $\phi$, in particular, if 
\begin{equation}
\label{Uqm}
\frac{\partial U(\phi)}{\partial\phi} =\frac{d\phi(\tau)}{d\tau}
+ \frac{\partial W(\phi)}{\partial\phi(\tau)}
\end{equation}
where $\tau\in\mathbb{R}$ and $W(\phi(\tau))$ is an ultralocal functional of
$\phi(\tau)$, we obtain the following stochastic equation for $\phi(\tau)$:
\begin{equation}
\label{langevin2}
\frac{d\phi(\tau)}{d\tau} = -\frac{\partial W}{\partial\phi(\tau)}+\eta(\tau)
\end{equation}
This is the Langevin equation that describes quantum mechanics, i.e. a quantum
field theory in one Euclidian dimension. The essential difference to
eq.~(\ref{langevin1}) is that we are not interested, only, in the limit
$\tau\to\infty$, but in  the solution for all values of $\tau$. This assertion
is meaningful only at the level of the correlation functions, of course. 

In this case, $\eta(\tau)$ is a Gaussian stochastic process, whose correlation
functions obey the same identities as in eq.~(\ref{gaussian_noise}), only the
time is, now, the Euclidian time. 

We are interested in the correlation functions,
$\left\langle\phi(\tau_1)\phi(\tau_2)\cdots\phi(\tau_n)\right\rangle$, of the
field $\phi$ and the identities that constrain them. 
We shall compute them from the following partition function 
\begin{equation}
\label{partition_function_phi}
\begin{array}{l}
\displaystyle
Z =
\int[d\eta(\tau) d\phi(\tau)]\,e^{-\int\,d\tau
\frac{\eta(\tau)^2}{2}}\,\delta\left(\eta(\tau)-\frac{d\phi}{d\tau}-\frac{\partial
  W}{\partial\phi(\tau)}\right) = \\
\displaystyle
\int[d\phi(\tau)]\,
e^{-\int\,d\tau\,
\frac{1}{2}\left(\frac{d\phi}{d\tau}+\frac{\partial
    W}{\partial\phi(\tau)}\right)^2}\,\left|\mathrm{det}\left(
\delta(\tau-\tau')\left(
\frac{d}{d\tau}
+ \frac{\partial^2 W}{\partial\phi(\tau)^2}\right)\right)\right| = \\
\displaystyle
\int[d\phi(\tau)d\psi(\tau)]\,
e^{-\int\,d\tau\,\left\{
\frac{1}{2}\left(\frac{d\phi}{d\tau}+\frac{\partial
    W}{\partial\phi(\tau)}\right)^2-\frac{1}{2}\int d\tau\,d\tau'\,
\psi_\alpha(\tau)\varepsilon^{\alpha\beta}\left(
\delta(\tau-\tau')\left(
\frac{d}{d\tau}
+ \frac{\partial^2 W}{\partial\phi(\tau)^2}\right)
\right)\psi_\beta(\tau')\right\}}
\end{array}
\end{equation}
We have introduced the determinant of the local operator,
$\partial^2U/\partial\phi(\tau)\partial\phi(\tau')$, in the action using
anticommuting fields, $\psi_\alpha(\tau)$. These are {\em not} ghosts! There
isn't any spin in one dimension, so the spin--statistics theorem is
vacuous~\cite{wightman}.

This is a formal expression, since the measure, $[d\phi(\tau)d\psi(\tau)]$, 
is not well-defined and the determinant of the operator is not well-defined 
either. Indeed, the passage from the first line to the second only holds, if
the determinant can't vanish--if it can, then, already, the first line is
ill-defined, since the delta function doesn't fully constrain the function
$\eta(\tau)$. So what we really are after is  a way to {\em define} the
expression in the third line--and use {\em that} definition to
define the expression in the first line.  We shall use the lattice. But, first, let us look at its symmetries.

\section{Elusive supersymmetry}
The action in eq.~(\ref{partition_function_phi}) is not invariant under transformations, that mix the commuting and the anticommuting variables.
 If we introduce, by hand, a second pair, $\chi_\alpha(\tau)$, of anticommuting variables, however,
 we may check that the action 
\begin{equation}
\label{S2fermion}
S=\int\,d\tau\,\left[-\frac{F^2}{2}+F\left(\frac{d\phi}{d\tau}+W'(\phi)\right)-\psi_\alpha\varepsilon^{\alpha\beta}\left(\frac{d}{d\tau}+W''(\phi)\right)\chi_\beta\right]
\end{equation}
is invariant, up to a total derivative, under the two transformations~\cite{dAFFV,damgaard_tsokos}:
\begin{equation}
\label{SUSY4}
\begin{array}{lcl}
\begin{array}{l}
\displaystyle
\delta_1\phi = -\zeta_\alpha\varepsilon^{\alpha\beta}\chi_\beta\\
\displaystyle
\delta_1\psi_\alpha = -\zeta_\alpha\left(\dot{\phi}+F\right)\\
\displaystyle
\delta_1 F = \delta_1\dot{\phi}\\
\displaystyle
\delta_1\chi_\alpha =0
\end{array}
&
\mathrm{and} &
\begin{array}{l}
\displaystyle
\delta_2\phi = \zeta_\alpha'\varepsilon^{\alpha\beta}\psi_\beta\\
\displaystyle
\delta_2\chi_\alpha = \zeta_\alpha'\left(-\dot{\phi}+F\right)\\
\displaystyle
\delta_2 F = \delta_2\dot{\phi}\\
\displaystyle
\delta_2\psi_\alpha =0
\end{array}
\end{array}
\end{equation}

These transformations are nilpotent, $\delta_A^2=0,A=1,2$ and do, indeed, close on the translations (in Euclidian time)~\cite{dAFFV,damgaard_tsokos}:
\begin{equation}
\label{SUSYQ}
\{\delta_1,\delta_2\}=-2\zeta_\alpha\varepsilon^{\alpha\beta}\zeta_\beta'\frac{\partial}{\partial\tau}
\end{equation}
They deserve, therefore, to be called supersymmetric. The target space of this theory is the real line. This supersymmetry is, of course, not ``target space'', but 
``worldline''. 

However, we notice that this action doesn't seem to stem from a Langevin equation. Also, that we've doubled the number of fermions and introduced a mismatch between fermions and bosons. In fact, one of the fermions is a gauge artifact, since only one appears through a derivative term and we may choose which one at any moment, there is a redundancy, i.e. a gauge symmetry. The gauge symmetry is ``$\kappa-$symmetry''~\cite{kappa_symmetry}. Furthermore, the fermionic kinetic term, in both actions, is a total derivative, thereby, providing only a surface term. The fermions,  thus, contribute ultra--local terms to the action, $\Psi_{\alpha,I}(\tau)K^{IJ}\varepsilon^{\alpha\beta}W''(\phi)\Psi_{\beta,J}(\tau)$, where $K^{IJ}$ is the ``bulk flavor mixing matrix'', that projects on the $I=1,I_3=1$ states of the two ``fermions''--and projects  out the $I=1,I_3=0$ and $I=0$ states (the ``boundary flavor mixing matrix'' projects on the two $I_3=0$ states). 
 The fermions may be, therefore,  integrated out exactly\footnote{Provided $W''(\phi)$ is of fixed sign. In ref.~\cite{nicolisSUSYQM} we discuss how to treat the case when $W''(\phi)$ can vanish in this case.}, 
 leaving a ``Polyakov loop'', the (path ordered) product of $\exp(\log W''(\phi))$, as an {\em ultra-local} contribution to the action, once we have gauge-fixed the partition function. The partition function in eq.~(\ref{partition_function_phi}) is the result of gauge-fixing the action in eq.~(\ref{S2fermion})~\cite{nicolisSUSYQM}. 
 We end up with a local, bosonic action:
 \begin{equation}
\label{Seffbos}
S[\phi]=\int\,d\tau\left[\frac{1}{2}\left(\frac{d\phi}{d\tau}+W'(\phi)\right)^2-\log W''(\phi)\right]=
\int\,d\tau\,\left[-\frac{1}{2}\phi\frac{d^2}{d\tau^2}\phi+\frac{1}{2}\left[W'(\phi)\right]^2-\log W''(\phi)\right]
\end{equation}
The observable we shall study is the auxiliary field, 
\begin{equation}
\label{auxF1}
F=\frac{d\phi}{d\tau} + W'(\phi)
\end{equation}
Supersymmetry implies that the auxiliary field has Gaussian correlation functions~\cite{stochastic,dAFFV,damgaard_tsokos}; in particular, vanishing 1--point function and an ultra--local propagator. These are the properties that we will try to check. We shall present results for the quartic superpotential, i.e. the anharmonic oscillator with ``sextic'' and ``quartic'' non-linearity:
\begin{equation}
\label{sexticV}
W(\phi)=\frac{m^2}{2}\phi^2+\frac{\lambda}{4!}\phi^4\Rightarrow W'(\phi)=m^2\phi+\frac{\lambda}{6}\phi^3\Rightarrow W''(\phi)=m^2+\frac{\lambda}{2}\phi^2
\end{equation}
with $m^2>0$ and $\lambda>0$.  Indeed, for one degree of freedom, if the potential is bounded from below, we can always write it as a perfect square.

\section{The auxiliary field on the lattice}
We discretize the action in the standard way and impose periodic boundary conditions on the scalar. The lattice action takes the following form
\begin{equation}
\label{Slatt}
S_\mathrm{latt}[\Phi]=\frac{1}{g m_\mathrm{latt}^2}\sum_{n=0}^{N-1}\left\{-\Phi_{n+1}\Phi_n+\Phi_n^2+\frac{m_\mathrm{latt}^4}{2}\left(\Phi_n+\frac{\Phi_n^3}{6}\right)^2-g^2 s m_\mathrm{latt}^4\log\left[\frac{1}{gs}\left(1+\frac{\Phi_n^2}{2}\right)\right]\right\}
\end{equation}
where we have introduced the lattice parameters
\begin{equation}
\label{lattparam}
\begin{array}{ccc}
\displaystyle
m_\mathrm{latt}^2=m^2a & 
\displaystyle \mathrm{and}&
\lambda_\mathrm{latt}=\lambda a^2 
\end{array}
\end{equation}
and the scaling parameters
\begin{equation}
\begin{array}{ccccc}
\displaystyle
\frac{\lambda_\mathrm{latt}}{m_\mathrm{latt}^4}=\frac{\lambda}{m^4}\equiv g, & &
\displaystyle 
a\frac{m_\mathrm{latt}^2}{\lambda_\mathrm{latt}}=\frac{m^2}{\lambda}\equiv s & \mathrm{and}&
\displaystyle
\Phi_n\equiv \phi_n\left(a\frac{m_\mathrm{latt}^2}{\lambda_\mathrm{latt}}\right)^{1/2}
\end{array}
\end{equation}
The scaling limit, therefore, consists in taking $m_\mathrm{latt}^2\to 0, \lambda_\mathrm{latt}\to 0$, keeping $g$ and $s$ fixed. If $g<1$ the theory is weakly coupled; if $g\geq 1$, it is strongly coupled. We remark that $g$ appears to be independent of the lattice spacing. 

The (rescaled) auxiliary field is given by the expression
\begin{equation}
\label{auxFlatt}
F_n=\frac{\Phi_{n+1}-\Phi_{n-1}}{2}+m_\mathrm{latt}^2\left(\Phi_n+\frac{\Phi_n^3}{6}\right)
\end{equation}
We see immediately that its 1--point function will vanish, $\langle F_n\rangle=0$, so we can use this as a check of the simulations. For the,  preliminary, results, presented below, we find $\langle F\rangle = 1.82689\times 10^{-5}\pm 1.66385\times 10^{-5}$, which is consistent with zero. We haven't studied the autocorrelation time here. 

\begin{figure}[thp]
\begin{center}
\includegraphics[scale=0.7]{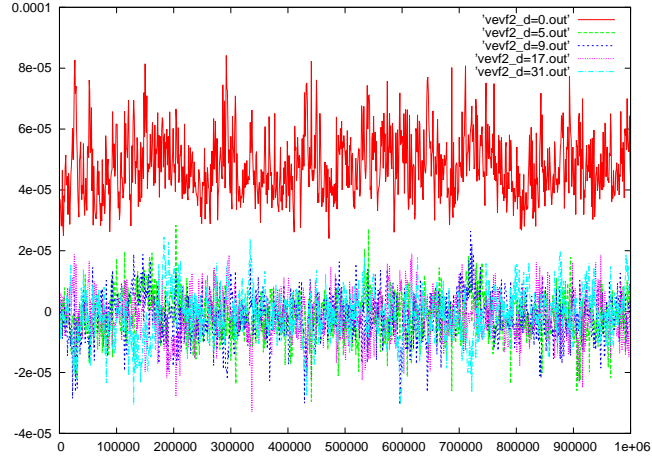}
\end{center}
\caption[]{Monte Carlo time series for the 2--point function for the auxiliary field, $\langle F_{|n-n'|}F_0\rangle$, for $|n-n'|=0,5,9,17,31$ for $N=64$. The red curve corresponds to $|n-n'|=0$, the others to $|n-n'|\neq 0$. Due to periodic boundary conditions, the relative distance cannot exceed $N/2$.}
\label{vevF2}
\end{figure}

\begin{figure}[thp]
\begin{center}
\includegraphics[scale=0.7]{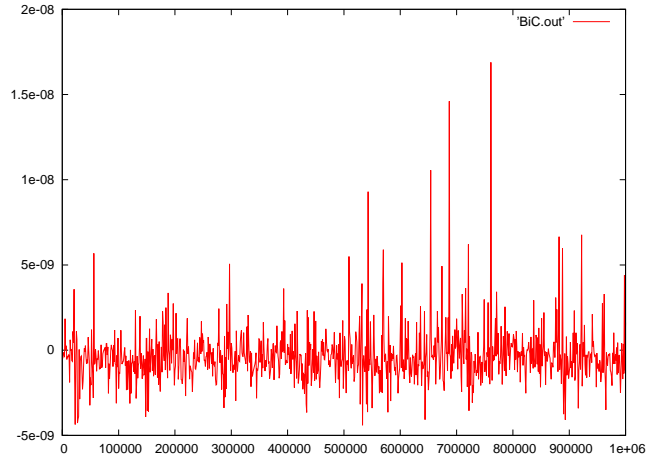}
\end{center}
\caption[]{Monte Carlo time series for the Binder cumulant for the auxiliary field, $\langle F^4\rangle-3\langle F^2\rangle$, for $N=64$.}
\label{BiC}
\end{figure}

The interesting quantities are the 2--point function and the ultra--local part of the connected 4--point function (the {\em Binder cumulant}). The former should be a $\delta-$function in the scaling limit and the latter should vanish to numerical precision. We display typical results of a Monte Carlo simulation  in figs.~\ref{vevF2}~and~\ref{BiC} 
\section{Conclusions}
We have studied quantum mechanics of a point particle in the stochastic formulation. Using the path integral formalism  the quantum fluctuations can be expressed locally through anticommuting variables. 
The physical import of these calculations is the following: the classical trajectory, $\phi(\tau)$, becomes a superfield $(\phi(\tau),\psi_\alpha(\tau),F(\tau))$.
The anticommuting variables enter the effective action in an ultra--local way, since their kinetic term is a total derivative, so the fermions are ``confined'' in bilinears.The effective action, in the stochastic formulation,  is the result of gauge fixing the $\kappa-$symmetry of a manifestly supersymmetric action.  It  is local and ``bosonic'' (depends only on commuting variables) and its correlation functions may be computed by standard numerical techniques. Supersymmetry is broken by the lattice, since the regularized auxiliary field does not have an ultra--local propagator, but is recovered in the scaling limit.

These results carry over to matrix quantum mechanics~\cite{matrixqm}, as well, since the Dirac algebra is, still, trivial. With a higher dimensional target space, however, it becomes possible to define a target space Dirac algebra, from the fermionic zeromodes~\cite{ABKW,banks_dixon}, and, thus, spinors and target space supersymmetry. This is, also,  how ``emergent spin''~\cite{creutz} might be realized. Details will be presented in future work.

{\bf Acknowledgements:} I'd like to acknowledge discussions with M. Axenides, C. Bachas, J. Bloch, E. G. Floratos, H. Giacomini, J. Iliopoulos,  C. Kounnas, 
G. Linardopoulos and T. Wettig.

\end{document}